\documentclass[useAMS,usenatbib]{mn2e}

\usepackage{natbib}
\usepackage{txfonts}
\bibpunct{(}{)}{;}{a}{}{,}
\usepackage[english]{babel}
\usepackage{ulem}

\usepackage{ifpdf}
\ifpdf 
\usepackage[pdftex]{graphicx}
\usepackage{epstopdf}
\else 
\usepackage{graphicx}
\fi

\newcommand{\be}{\begin{eqnarray}}
\newcommand{\ee}{\end{eqnarray}}

\def\figsize{3.3in}

\usepackage{colordvi}

\RequirePackage[colorlinks=true
,urlcolor=blue
,anchorcolor=blue
,citecolor=blue
,filecolor=blue
,linkcolor=blue
,menucolor=blue
,pagecolor=blue
,linktocpage=true
,pdfproducer=medialab
]{hyperref}

\begin{document}

\title{The Fate of a WD Accreting H-Rich Material at High Rates}

\author[Idan et al.]{Irit Idan$^1$, 
Nir J. Shaviv$^2$ and Giora Shaviv$^1$\\
\noindent
$^{1}$ Dept. of Phys., Israel Institute of Technology, Haifa 32000, Israel \\
\noindent
$^{2}$ Racah Institute of Physics, The Hebrew University,  Jerusalem 91904, Israel}

\pagerange{\pageref{firstpage}--\pageref{lastpage}} \pubyear{2010}

\maketitle

\begin{abstract}
We study C/O white dwarfs with masses of 1.0 to 1.4 $M_\odot$  accreting solar-composition material at very high accretion rates. We address the secular changes in the WDs, and in particular, the question whether accretion and the thermonuclear runaways result is net accretion or erosion. The present calculation is unique in that it follows a large number of cycles, thus revealing the secular evolution of the WD system.  

We  find that counter to previous studies, accretion does not give rise to steady state burning. Instead, it produces cyclic thermonuclear runaways of two types. During most of the evolution, many small cycles of hydrogen ignition and burning build a helium layer over the surface of the white dwarf. This He layer gradually thickens and progressively becomes more degenerate. Once a sufficient amount of He has accumulated, several very large helium burning flashes take place and expel the accreted envelope, leaving no net mass accumulation. 
 
The results imply that such a system will not undergo an accretion induced collapse, nor will it lead to a SN Type Ia, unless a major new physical process is found. 
\end{abstract}


\section{Introduction}

The  prevailing scenarios leading to Type Ia SNe can be divided into two classes - those of singly (SD) and those of doubly degenerate (DD) systems. In the first, a degenerate WD accretes from its binary companion, and accumulates sufficient mass to approach the Chandrasekhar limit, where CO detonation or the collapse to a neutron star takes place \citep[e.g.,][]{Hillebrandt2000,Pod2008}. In the second class, Ia's occur from the merger of a binary WD system \citep[e.g.,][]{Webbink1984}. Since no model within the above classes of progenitors comes without significant caveats, there is still no consensus model for Type Ia SNe.

When considering the merger of WDs, one major caveat is the merger rate. The theoretical and observed rates of super-Chandrasekhar mergers is typically an order of magnitude smaller than the observed rate of Ia's, unless carbon detonation can take place in sub-Chandrasekhar systems as well  \citep{vanKerkwijk2010,Badenes2012}.  Another problem is that merger scenarios would tend to produce explosions which are more heterogeneous than observed Ia's, both in terms of light curves and ejected mix of elements. 

Systems where a WD accretes from a non-degenerate companion, on the other hand, are subject to other theoretical and observational limitations. Perhaps the most important one is that of the accretion rate. In their review, \cite{Hillebrandt2000} pointed out that although SD is the favoured progenitor model, the major problem has always been that nearly all possible accretion rates can be ruled out by strong arguments. 

When hydrogen is accreted at sufficiently low rates, it can cool and become degenerate. When a sufficient amount of gas accumulates, the hydrogen ignites to produce an unstable thermal flash \citep{Schwarzschild1965,Rakavy1968}. Since the nuclear luminosity may reach extremely high values, of the order of $10^{12}L_{\odot}$ for a relatively long period of time, the flash can dynamically eject the accreted mass. In fact, this mechanism naturally explains classical novae eruptions \citep{Iben1984}. Because lower accretion rates allow more mass to accumulate, lower accretion rates produce stronger eruptions.  Moreover, the observations indicate that the ejecta contain as a rule, more heavy elements than in the accreted matter.  It is obvious that if He burning and beyond does not take place, then these heavy element originate form the underlying WD, which is eroded in this process. 

\cite{Prialnik1995} and \cite{Yaron2005} found that all WD's accreting at rates  ${\dot m}\lesssim 10^{-7}M_{\odot}/yr$ erode in mass. At somewhat higher accretion rates, there is net accumulation, but with a low efficiency since most of the mass is still ejected. 

One way to avoid the hydrogen flashes associated with the low accretion rate, and the consequent mass loss, is to accrete pure helium, which may avoid ignition \citep{Iben1991, Iben1994}. However, population syntheses suggest that such progenitor systems, which appear as AM CVn stars can
contribute at most, about $10^{-2}$ of the SN Ia rate \citep{HeAccretionPopulation}. 

At the opposite limit, of very high accretion rates, \cite{Iben1984} argued that the WD does not experience hydrogen shell flashes, but instead, its envelope expands to a radius $R\approx 1050(M/M_{\odot}-0.5)^{0.68}R_{\odot}$, which for a massive WD is much larger than the orbital separation of typical cataclysmic variables. Once a common envelope is formed, heavy mass loss should then prevent the mass of the C/O WD from reaching the Chandrasekhar limit. 

In between the above ranges, are accretion rates for which the released gravitational binding energy is close to the Eddington limit, around $10^{-7}$ to $10^{-6}M_{\odot}/yr$ (depending on the WD mass). It was suggested that the high accretion should lead to a quiet hydrostatic burning of H and He \citep{Nomoto1982}.

Various attempts to calculate the effect of high accretion rate were carried out under different approximations (e.g., \citealt{Nomoto1982}, using the technique developed by \citealt{Nomoto1977}). In particular, the assumption of steady state ignores the time dependent evolution of the accretion system. However, this assumption simplifies the calculation because time dependent models have two major obstacles. First, the latter require very fine mass shells, and second, the full evolution requires following a large number of small flashes. 

\cite{Fujimoto1982} investigated the thermal properties of hydrogen shell burning on accreting white dwarfs. Assuming hydrostatic equilibrium (no dynamic effects were allowed),
Fujimoto found that the hydrogen burns steadily on the surface of the WD. The gravitational energy release was simulated by 
$${\partial \over \partial q}\left({L_r \over M}\right)=\epsilon_g,$$ where $L_r$ is the radiative flux and $q=M_r/M$.  The burning shell exhibited periodic flashes, all treated under hydrostatic equilibrium. As Fujimoto assumed steady state, the rate at which the mass of the  WD grew was equal to the assumed accretion rate.  

\cite{Prialnik1995} alleviated the assumption of steady state. The have shown that one must calculate at least several dozens of cycles before the effects of the arbitrary chosen initial conditions decays away. In some cases even 1001 cycles were followed  \citep{Epel2007}. The calculations were extended by \cite{Yaron2005} with basically the same result. In particular, it was found that WDs accreting at rates  ${\dot m}\gtrsim 10^{-7}M_{\odot}/yr$ grow in mass.

In addition to the above theoretical arguments, interesting observational constraints should be considered as well. Some argue in favour of SD scenarios and some argue against.  

The SD scenarios imply that the former companion stars should remain after the explosion, and be observable in association with nearby remnants. And indeed, a G-star was claimed to be the former companion of Tycho Brahe's 1572 supernova \citep{ruiz2004binary}. On the other hand, the opposite claim was also made. No former companion was found to be associated with SNR 0509-67.5, down to $M_v = + 8.4$ \citep{schaefer2012absence}. In this respect, one should also mention the lack of any radio detection in SN 2011fe, which took place only 8 Mpc away. This was used, under several theoretical assumptions, to place an upper limit of order 10$^{-8}$M${}_{\odot}$/yr, on the companion wind prior to the supernova explosion \citep{Horesh}. Using {\sc Chandra} and {\sc HST} archival  data, SN 2011fe could also be used to place a limit on the accretion spectrum. If the WD was emitting at the Eddington luminosity, then its temperature should have been within $60$ eV $\gtrsim kT \gtrsim 10$ eV \citep{Liu2012}.  In contrast, \cite{Voss2008} reported the disappearance of an X-ray source from the location of a Type Ia supernova in NGC 1404. They estimated the source to have been emitting  $2-3 \times 10^{37}$ erg/sec in X-rays.

In terms of statistics, high rate accretion implies that SD progenitor systems should be sufficiently bright as to be easily detected. It is presently not clear at all, whether SD progenitors can be identified with any of the known cataclysmic binaries, the known symbiotic systems, or whether perhaps with the very bright supersoft X-ray sources \citep[][]{Iben1994,Gilfanov2010}.

Given the present state of uncertainty, further study of the different progenitor scenarios is necessary. Here we shall address one particularly pressing question, which is whether systems accreting hydrogen at high rates can actually accumulate mass.

We begin in \S\ref{sec:Details} by describing our  numerical code, and the delicate points requiring attention when carrying out a multi-cycle evolution. In \S\ref{sec:results} we describe our results. Here we will see that there are two types of thermonuclear runaways (TNRs), those which burn H and expel no matter, and larger ones burning He, which do do expel. We will therefore discuss in \S\ref{sec:energetics} the energetics of how this is possible, even though the specific energy release in helium burning is small. We end with conclusions in \S\ref{sec:conclusions}.

\section{Details of calculations}
\label{sec:Details}

Our main purpose is to investigate the long term secular behaviour of  high accretion rate systems. We know from the work of \cite{Kovetz1994} that a relatively large number of cycles is required for the detection of the small secular effects. For this reason, we ensured that the code can simulate thousands of cycles. Even so, the calculation was carried out without compromising the physics. 

The initial models included WDs with masses of  $1$, $1.25$, $1.35$ and $1.4M_{\odot}$,  despite the findings of \cite{Weidemann2000} that this is too high and may be very rare. The models are of  bare C/O WDs, namely, there is no atmosphere composed of lighter elements, such that the C/O composition extends  to the surface. The initial core temperature is between  $10^7$K and $5\times 10^7$K. However, as we will show later,  the final results are insensitive to the core temperature. In addition, the models assume no rotation of the WD, and that the dissipation of the rotational energy at the boundary layer is complete such that the entire rotational energy is radiated away. These conditions are  the most favourable for obtaining highly degenerate matter at the surface.

All models considered accrete solar composition material at a rate of $10^{-6}M_{\odot}/yr$. 
According to \cite{Nomoto1982}, this accretion rate is large enough to puff the accreting system into a red-giant-like-star burning at steady state, with a slow evolution on a timescale determined by the nuclear burning shell. 

 The cyclic phenomenon demands very fine division during mass loss and peak burning, as a coarse discretisation was found to introduce unacceptable errors. Therefore, the minimal mass shell was chosen to be between  $10^{-7}-10^{-9}M_{\odot}$, depending on the mass of the WD. At the peak, over 8000 mass shells were integrated. 

The code we use is based on \cite{prialnik1981,prialnik1986}. It includes diffusion, convection, and all relevant nuclear reactions up to A=40, which consider the intermediate electrostatic screening correction. 

To a large extent, mass loss following a TNR can occur through two types of mechanisms. The first type is through an ``explosion". Namely, a large release of energy can be used to eject mass until the star settles into a dynamic steady state. However, after a few dynamical times have elapsed, a steady state wind can be driven through a continuum driven wind \citep{Bath1976}. The numerical code should be able to describe both types of mass loss. This can be done by employing a general mass loss condition. 

The ejection of a mass element occurs, irrespective of the mechanism, if its velocity is larger than both the speed of sound and the escape velocity at 
the radius from which the mass is removed \citep{Prialnik1978,Prialnik1979}. This general rule leads to very extended radii and short time steps before the conditions are satisfied. \cite{Kato1983} was the first to fit a steady state wind in a model with two parameters which were the mass loss rate and the mass of the envelope. The envelope was assumed to be in place as initial conditions. Kato found that steady mass-loss solutions can be found only if the luminosity at the bottom of the envelope is greater then the minimum value of the local Eddington luminosity throughout the envelope. 

Here we follow \cite{Kovetz1998}, and check at every time step whether an optically thick wind can exist, and remove mass accordingly. We stress that in no case, did the removal of the mass was found to induce additional mass removal, namely, no rarefaction wave was created leading to endless mass loss.  The possibility of an optically thin wind was not considered in this calculation. 

We quench the accretion during the eruption, because we expect the accretion disk to be destroyed under the dynamic and high luminosity conditions. Thus, mass accretion is switched off when the nuclear luminosity is larger than 10 times the total luminosity, and switched on again when the luminosity decreases below the threshold\footnote{Using other conditions, such as switching off the accretion only when mass loss is present gave rise to minor changes in the time scale.}. For this reason, the long term average accretion is typically a third of the peak instantaneous accretion rate of $10^{-6}M_{\odot}/yr$.

\section{Results}
\label{sec:results}

\begin{figure}
\centering
\includegraphics[width=\figsize]{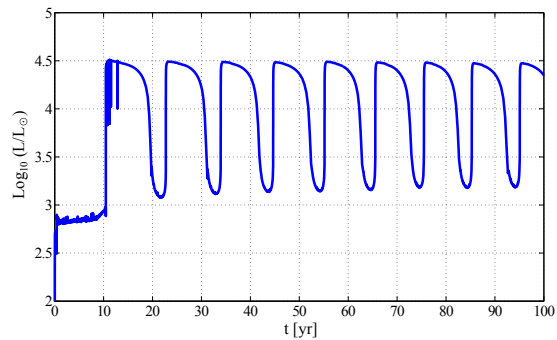}
\caption{The first 10 outbursts for a $1.0 M_\odot$ WD with peak accretion rate of $10^{-6}M_\odot yr^{-1}$ -- only the first outburst ejected mass.}
 \label{fig:fig1}
\end{figure}

Figure\ \ref{fig:fig1} depicts the initial behaviour of the $1.0 M_\odot$ WD model, which settles into steady cycles of small outbursts without ejecting any mass. Unlike accretion at low rates, giving rise to classical or recurrent novae \citep{Rakavy1968}, the high rate accretion keeps the accreted layers barely degenerate. As a consequence, the TNR is modest and involves a small shell. 

\begin{figure} 
\centering
A\includegraphics[width=\figsize]{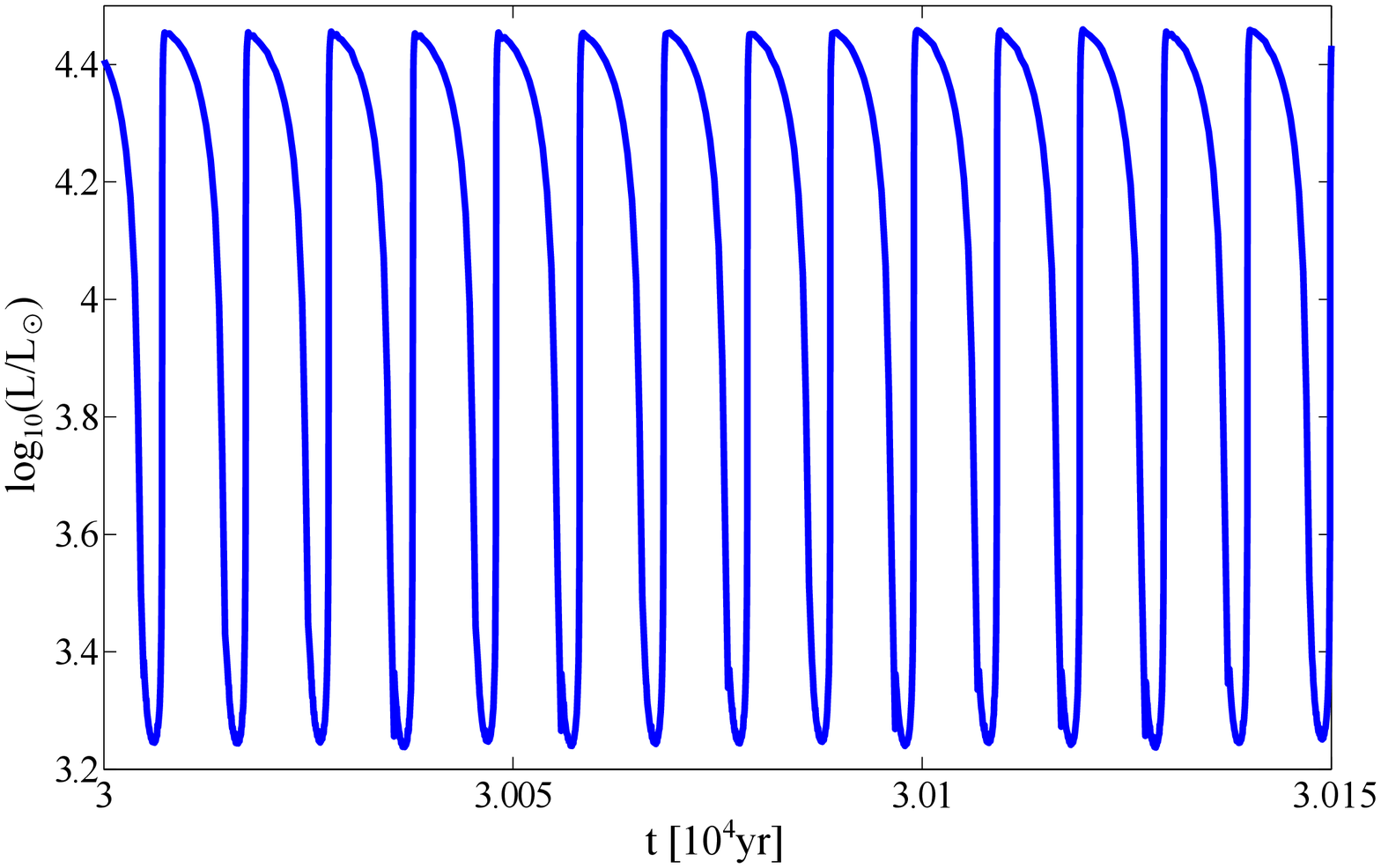}
B\includegraphics[width=\figsize]{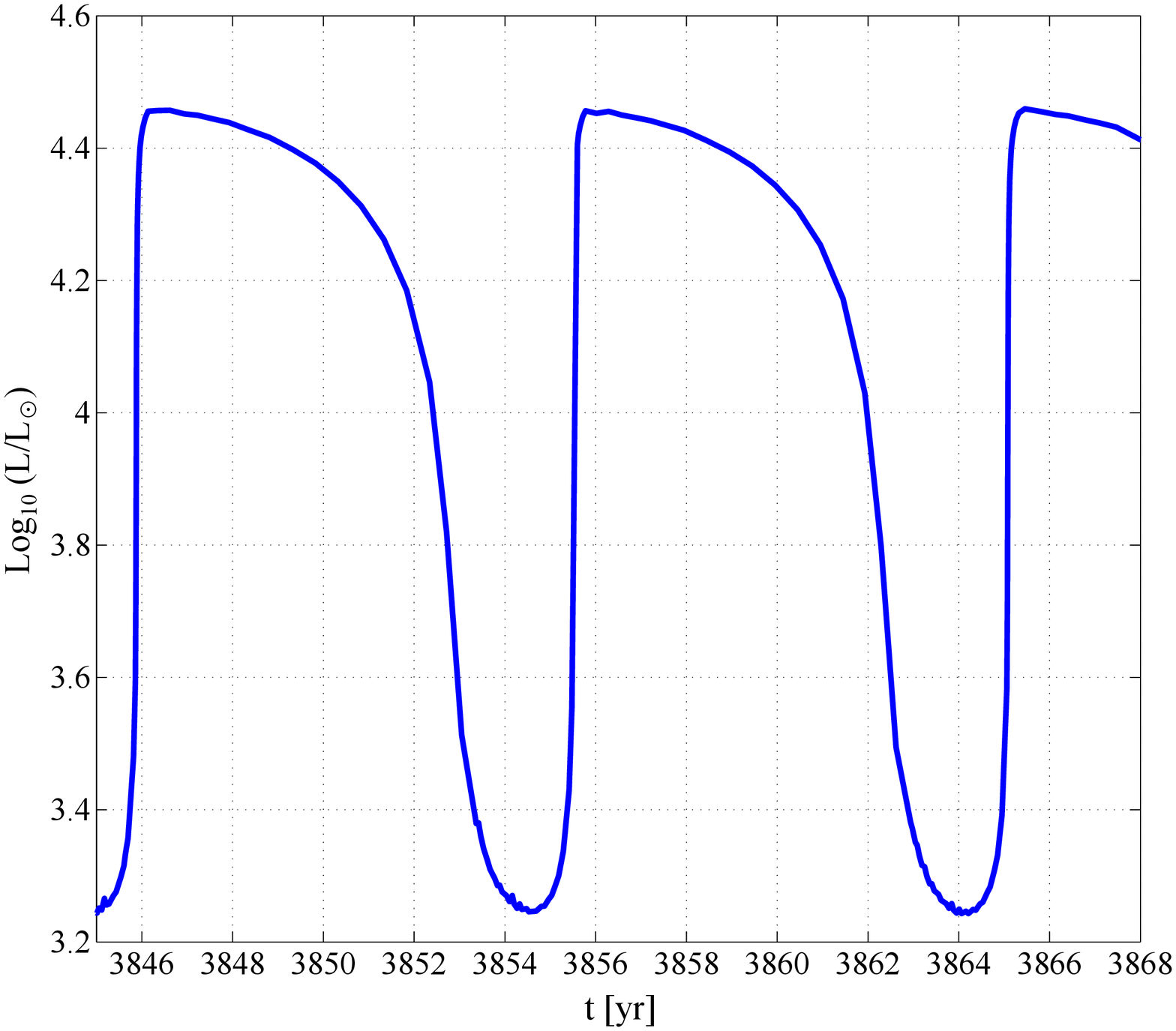}
 \caption{The secular sequence of flashes for the model depicted in fig.\ \ref{fig:fig1}, $3 \times 10^4$ years after the onset of accretion. Panel B shows one cycle in detail. Note the high similarity from cycle to cycle.  }
 \label{fig:fig2}
\end{figure}

Figure\ \ref{fig:fig2} demonstrates that the cyclic behaviour continues for many cycles, with a period of about 10 yr. There is very little secular change from one cycle to the next. In each cycle, hydrogen undergoes a small runaway in which the luminosity increases by over a magnitude. The flashes are characterised by a very sharp rise and a slower decline. A refined picture is shown in panel B of fig.\ \ref{fig:fig2}, demonstrating how the system approaches the Eddington luminosity for a couple of years. The system stays in the  minimum state only a very short fraction of the total period, until thermal ignition starts a new cycle.


\begin{figure}  
\centering
\includegraphics[width=\figsize]{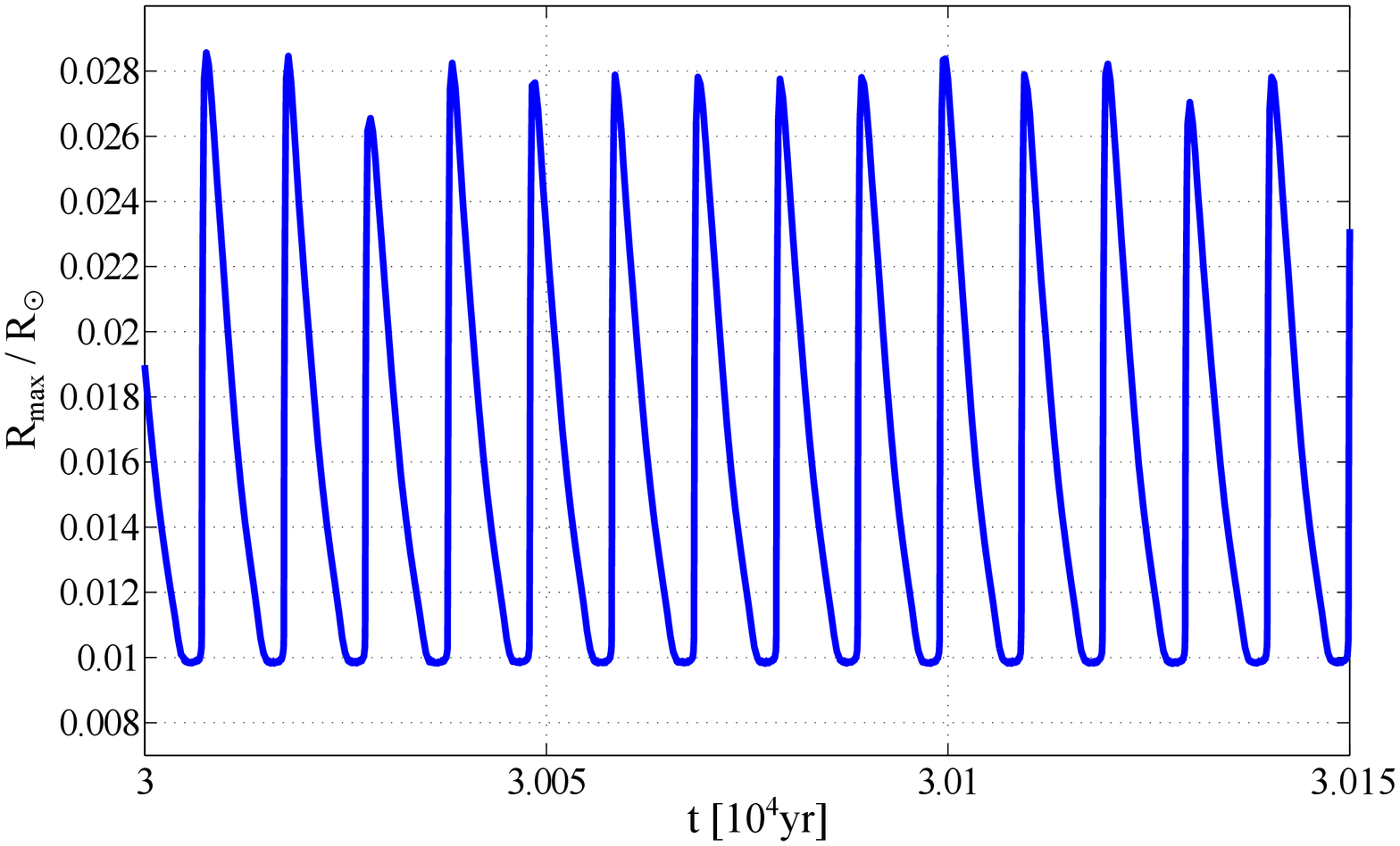}

 \caption{The maximum radius reached during outbursts, for the model depicted in fig. \ref{fig:fig1}. The small variations in the maximum radius are due to the finite time step taken.}
 \label{fig:radius}
\end{figure}

Because the emitted luminosity during the flash is slightly less than the Eddington luminosity, no mass loss develops. This is because the luminosity was used to expand the thin hydrogen shell (see fig.\ \ref{fig:radius}).

\begin{figure}  
\centering
\includegraphics[width=\figsize]{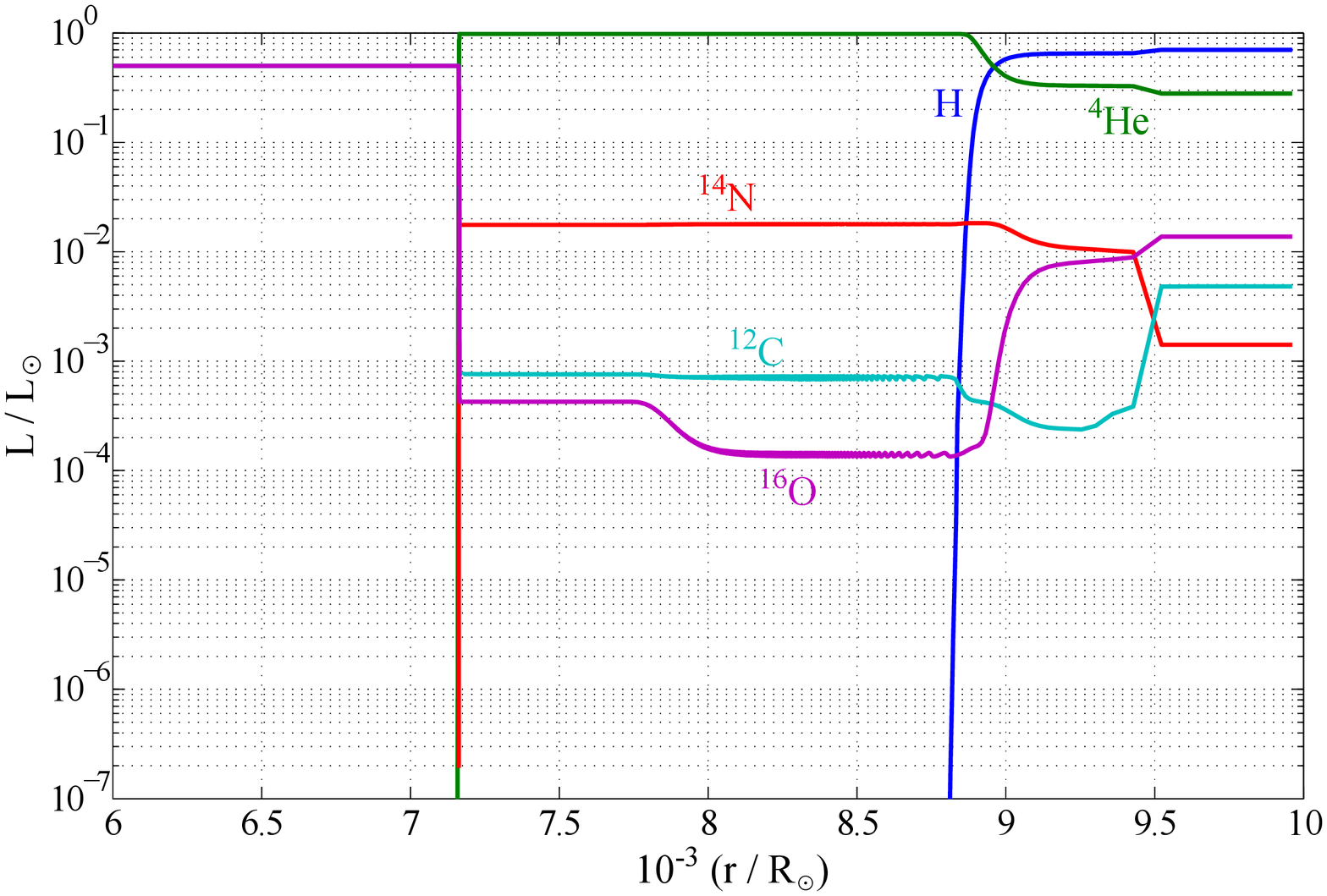}

 \caption{The abundances after 3000 cycles of small outbursts, for the model depicted in fig. \ref{fig:fig1}.}
 \label{fig:abundances}
\end{figure}

The burnt hydrogen accumulates as helium. The abundances profile after 3000 cycles of small outbursts are shown in fig.\ \ref{fig:abundances}. The underlying CO WD is present at small radii, below $\sim 7\times 10^{-3} R_\odot$. It is surrounded by a thick layer of helium with smaller amounts of C and O. The  freshly accreted material having solar composition is located at the outer radii.  It is important to note that the burning is almost complete. At this phase the system appears as a perfect recurrent object which accretes the ashes onto the original WD.

\subsection{Secular changes in the accreting star}

\begin{figure} 
\centering
\includegraphics[width=\figsize]{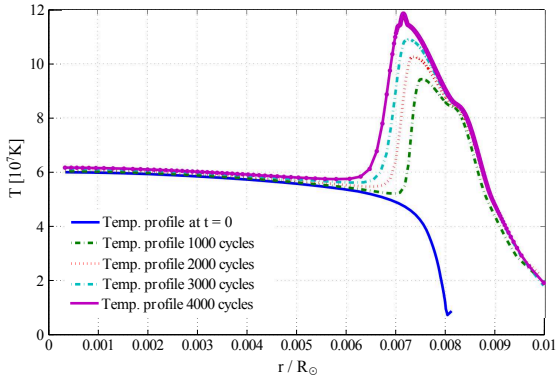}

 \caption{Temperature profiles at different cycles, for the model depicted in fig. \ref{fig:fig1}. The slow diffusion of heat into the WD is due to the temperature inversion which inhibits convection, and the fast time scales associated with the high accretion rate which limits the energy lost outwards. }
 \label{fig:comparison1}
\end{figure}

The initial profile which we employ is obtained by evolving a polytropic WD, until a desired central temperature is reached. Once accretion cycles begin, the small hydrogen flashes give rise to a temperature wave which propagates into the helium layer and the outer edge of the CO WD. This wave increases the temperature at the base of the accreted helium envelope to nearly  $1.2\times 10^8K$ after 4000 cycles. Fig.\ \ref{fig:comparison1} provides snapshots of the temperature profile. 

\begin{figure}  
\centering
\includegraphics[width=\figsize]{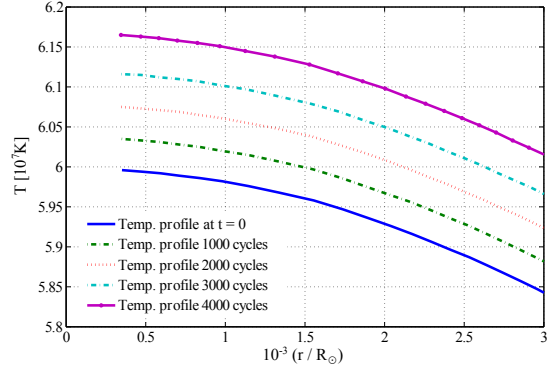}

 \caption{The temperature profile at the core, for the model depicted in fig. \ref{fig:fig1}. Note that the heating is due to the adiabatic contraction of the WD associated with the additional accreted mass.}
 \label{fig:coreT}
\end{figure}

The large mass of the accreted envelope, nearly $0.01M_\odot$, affects the structure of the WD as well. As a consequence, the WD contracts and the inner temperature rises from  $6.0 \times 10^7K$ to $6.17\times 10^7K$---a $3\%$ change (see fig.\ \ref{fig:coreT}).

\begin{figure}  
\centering
\includegraphics[width=\figsize]{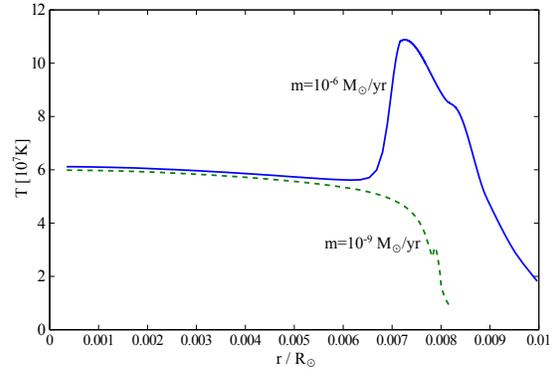}
 \caption{Comparison between the temperature profiles after 3000 outbursts of a 1 $M_\odot$ WD, accreting at a rate of $10^{-9} M_\odot/yr$ (dashed)  and at a rate of $10^{-6}M_\odot/yr$ (solid). }
 \label{fig:comparison}
\end{figure}

It is interesting to compare the temperature profiles of the WD for different accretion rates.  We therefore, calculated using the same initial WD structure a model having an accretion rate of $10^{-9} M_\odot yr^{-1}$. We follow the calculation for 3000 cycles. Note that in both case, this corresponds to a total accreted mass of order $10^{-2} M_\odot$. See fig.\ \ref{fig:comparison}.

For the low accretion rate, where the accreted material is sufficiently degenerate, the outbursts are strong enough to eject all the accreted material, as well as some of the eroded WD. The interval between outbursts ($3\times 10^4 yr$) is long, such that the WD can cool down. The temperature profile of the WD accreting is then hardly affected by the outburst itself. The only evidence for the existence of the previous outbursts is a small hump in temperature profile.

The outcome of the high accretion rate is completely different. Since the material is insufficiently degenerate, the pressure at the base of the envelope prior to the outburst,  $P_{crit} \approx GM_{wd} m_{env} / 4\pi R^4_{wd} \approx 8-9 \times 10^{17} dyn \ cm^{-2}$, is too low to allow mass ejection. This result is in agreement with the analytical calculation of \cite{Fujimoto1982} and \cite{MacDonald1983} that predicted mild outbursts occur for $P_{crit} \lesssim 10^{18} dyn \ cm^{-2}$. Because the outburst is weak, matter cannot escape the gravitational field of the WD. Consequently, a heat wave can form and penetrate into the base of the accreted envelope. This heat wave will serve as the trigger for the giant outburst described in the next section. 

To summarise, the positive temperature gradient in the high accretion rate inhibits heat flow into the core. The degree of heating per cycle is negligible, but accumulates during thousands of cycles. 

\subsection{The final outburst}

\begin{figure} 
\centering
\includegraphics[width=\figsize]{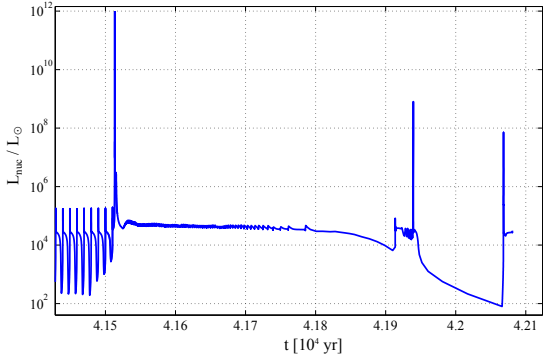}

 \caption{The nuclear luminosity  during the final outbursts.}
 \label{fig:flashes4}
\end{figure}

The surprise came after 4153 cycles when the accreted mass reached $10^{-2}M_{\odot}$. The maximum temperature in the helium layer reached $1.2\times 10^8$K, allowing helium ignition in a strong flash (see figure\ \ref{fig:flashes4}). 

The peak nuclear energy generation during the He TNR reached $10^{12}$erg/sec  and the maximum temperature reached a peak of $4\times 10^9$K. 

The very strong He flash caused a temperature wave which propagated to the hydrogen burning shell located above the pure helium layer. The wave raised the temperature, and accelerated the burning accordingly. This resulted in the ejection of about 1/3 of the helium layer accreted. This phenomenon repeated two more times with the net outcome that  the entire accreted helium layer was ejected.

These results should be compared with those of \cite{Yaron2005}, who calculated a similar model. 
For $\dot{m}_{acc}=10^{-8}M_{\odot}/yr$, they have found that the WD erodes. For $\dot{m}_{acc}=10^{-7}M_{\odot}/yr$, the white dwarf gained mass through outbursts, while for $\dot{m}_{acc}=10^{-6}M_{\odot}/yr$ it gained mass without outbursts. However, they only followed a limited number of cycles, which did not allow for He ignition. 

 We repeated the calculation for colder WDs, having lower core temperatures. We found minor changes. For example, the time between outbursts increased from 8 yr for a $ 6\times 10^7{\rm K}$ WD to 9.5 years when the central temperature was $10^7{\rm K}$. He ignition took place after 4900 accretion cycles, corresponding to an accreted mass of $0.0103M_\odot$.
 
We also repeated the calculation for WDs having different masses.   
 The time between outbursts varied from 8 years for a $1M_\odot$ to three months for the $1.35 M_\odot$. But the main results remain the same.  The results are summarised in  Table\ \ref{tab:mass}. In all three calculations, the last stage was nearly the same---once the pressure at the base of the He envelope became sufficiently high, the He ignited and the entire accreted envelope was ejected. 

Unlike lower mass WDs, the $1.4M_\odot$ WD behaved differently. During the initial hundred or so cycles, it followed a similar behaviour as lower mass WDs---small outbursts took place with a repetition period of $\sim 18$ days.  However, after roughly 100 cycles, the WD expand and became a red giant. The initial cyclic behaviour we find is consistent with \cite{Kovetz1994}, who investigated the effect of accretion onto a  $1.4M_\odot$ CO WD at different accretion rates, between $10^{-10}$ to $10^{-6}M_\odot/yr$. 

\begin{table}
\caption{The total accumulated mass as function of WD mass for $\dot M=10^{-6}M_\odot yr^{-1}$ before completely ejected by the final outburst}
\vskip -5mm
\label{finalm}
\begin{center}
\begin{tabular}{cc}
\hline
WD Mass [$M_\odot$]  & Final accreted mass [$M_\odot$] \\
\hline
1.0  &$ 9.9\times 10^{-3}$ \cr
1.25  &$ 7.5\times 10^{-4}$ \cr
1.35  &$ 1.1\times 10^{-4} $
\\
\hline
\end{tabular}
\end{center}
\label{tab:mass}
\end{table}

\subsection{Abundances}

The abundances in the mass ejected in the final three outbursts are given in Table\ \ref{tab:abund}. We find the following. 

\begin{enumerate}
\item There is very little hydrogen ejected. This is because most of the hydrogen was converted into helium during the small outbursts. The small amount which is ejected is from the outer most accreted layer which did not have time to burn. 

\item Helium abundance increased to about 0.7. This is more than twice the solar composition. However, it is less than the initial X+Y, because of the triple $\alpha$ reaction. 

\item The Z increased by more than an order of magnitude, from 0.02 to nearly 0.29. The interesting point, however, is that the isotopic ratios are different from those in the standard solar composition (or different for standard nova ejecta). 

\end{enumerate}

\begin{table}
\caption{Ejecta Abundances.}
\begin{center}
\begin{tabular}{|c|c|c|}
\hline
Element  & Solar&Maximal mass \cr
\hline
H & 0.7&6.3(-5 )\cr
$^{4}{\rm He}$  &0.28&  0.71  \cr
$^{12}{\rm C}$ & 3.9(-3)& 1.25(-1) \cr
$^{14}{\rm N}$ &1.0(-3)&5.4(-3) \cr
$^{16}{\rm O}$ & 9.4(-3)&5.7(-2)  \cr
$^{20}{\rm Ne}$&0.0  &4.3(-2)  \cr
$^{23}{\rm Na}$&0.0  &1.5(-4)  \cr
$^{24}{\rm Mg}$&0.0 &5.9(-2) \cr
$^{27}{\rm Al}$ &0.0 &1.2(-5) \cr
$^{28}{\rm Si}$ &0.0 &3.8(-4)  \cr
\hline
\end{tabular}
\end{center}
\label{tab:abund}
\end{table}%

\subsection{Understanding the energetics}
\label{sec:energetics}
A simple analytic approximation for the radius of a zero temperature WD was given by \cite{Verbunt1988}, it is
\begin{equation}
{R_{wd}\over R_{sun}}\approx  0.0114 { \left( \left({M_{wd}/ M_{ch}}\right)^{-2/3}-\left({M_{wd}/ M_{ch}}\right)^{2/3} \right)^{1/2}  
\over \left(1+3.5\left({M_{wd}/ M_p}\right)^{-2/3}+\left({M_{wd}/ M_p}\right)^{-1}\right)^{2/3} },
\label{eq:Rmass}
\end{equation}
where $M_p = 0.00057 M_\odot$ is a numerical constant. Using this expression, we can calculate the critical mass $M_{crit}$ above which the binding energy of an element is larger than the specific energy $\epsilon$ released in nuclear burning. Namely, we require that 
\begin{equation}
\epsilon = {GM_{crit}\over R_{wd}}<E_{bind},
\label{eq:condition}
\end{equation}
where $E_{bind}$ is the nuclear binding energy per unit mass. 
The results are given in table \ref{tab:crit}.  

According to the standard lore, hydrogen burning can therefore eject mass from WDs having masses of up to $1.436M_{\odot}$. On the other hand, helium burning can eject mass provided the mass of the WD is less then $1.3M_{\odot}$.  Nevertheless, this is the opposite of what the present models exhibit. 

In all our $10^{-6} M_\odot/yr$ models, with WD masses up to 1.35$M_\odot$, the hydrogen flashes did not expel any material even though their mass was below the critical one. On the other hand, all these models did expel the helium accumulated from the hydrogen burning, even though the most massive of which were already above the critical WD mass for helium ejection.

\begin{table}
\caption{The critical mass $M_{crit}$ above which the binding energy is greater than the energy released in nuclear burning.}
\begin{center}
\begin{tabular}{|c|c|c|cl}
\hline
Reaction  & Energy release [MeV]  & Q [erg/gr] & $M_{crit}$  \cr
\hline
$4p\rightarrow {\rm He} $  & 26.47 & $6.3\times 10^{18}$& 1.436 \cr
$3{\rm He} \rightarrow {\rm C }$  & 7.276 & $5.8\times 10^{17}$ &1.30  \cr
${\rm C+C} \rightarrow {\rm Mg}$ & 7.075 & $2.8\times 10^{17}$ & 1.29  \cr
\hline
\end{tabular}
\end{center}
\label{tab:crit}
\end{table}%

To understand how hydrogen burning could result with no expulsion, but He can, let us consider the energetics. 

When hydrogen is initially accreted on the bare WD, at an average rate $\dot{m}$, it releases its binding energy $ (G M / R_{WD}) \dot{m}$. In addition, as it burns, it releases $\dot{m} \epsilon_H$, where $\epsilon_H$ is the specific nuclear energy released in hydrogen burning. On the other hand, the maximal luminosity with which the system can lose energy is $ \eta L_{Edd}$, with $\eta$ being some factor smaller than unity. This factor depends on the actual luminosity of the expanded state, and its duty cycle. For the $1 M_\odot$ WD accreting at $10^{-6} M_\odot/yr$, it is $\eta \sim 0.5$.  

Comparing these two energies, gives $ \dot{m} \left( {G M / R_{WD}} + \epsilon_H\right) > \eta L_{Edd}$.
In other words, every H burning cycle releases more energy than can be radiated away. As a consequence, the envelope must expand to some effective radius, for which  the specific binding energy satisfies 
\begin{equation}
\dot{m} \left( {G M\over R_\mathrm{eff}}  - u_{int}  + \epsilon_H \right) \approx \eta L_\mathrm{Edd}.
\label{eq:energy}
\end{equation}
Moreover, the gas in the expanded state is hot, and therefore its specific internal energy $u_{int}$ is important, and can be as much as half the gravitational binding energy. 

Once helium is ignited, the specific binding energy of the puffed up envelope is therefore smaller than the specific binding energy it would have had if the helium in the envelope was degenerate and located at $r \approx R_{WD}$. Thus, when helium ignition takes place, the specific nuclear energy released is sufficient to expel the envelope. Namely, 
\begin{equation}
  - {G M\over R_\mathrm{eff}} + u_{int} + \epsilon_{He} > 0.
 \label{eq:He}
\end{equation}
Clearly then, the main reason why the helium burning can expel matter is because the high accretion rate forced the accreted matter to expand and heat such that it had a smaller specific binding energy.
One can recast this differently. If we look at the average binding energy, $-GM/R_\mathrm{eff}+ u_{int}$  from eq.~\ref{eq:energy}, and plug it in eq.~\ref{eq:He}, we find that the condition for expulsion is:
\begin{equation}
\epsilon_H + \epsilon_{He} > {\eta L_\mathrm{Edd} \over \dot{m}}.
 \label{eq:expul}
\end{equation}
Thus, the high accretion rate, the hydrogen burning, and the limited luminosity combine to keep the envelope bloated enough to allow expulsion. However, an accretion rate which is too high would force the system to become red-giant-like, and the envelope would be expelled. 

Note also that eq.~\ref{eq:energy} should include two additional components.  First, some of the heat in the envelope should diffuse into the WD. This implies that there is an additional thermal energy term which should in principle appear. However, when this energy was artificially omitted, the final results hardly changed. 

Second, an additional energy gain by the system is associated with the increased binding energy of the underlying WD---the latter contacts and adiabatically heats.  However, since the heat cannot diffuse out over the $n$ accretion cycles,  it is stored and released upon the final ejection. In other words, this energy component is unimportant.

\subsection{Few more results}
\begin{figure}  
\includegraphics[width=\figsize]{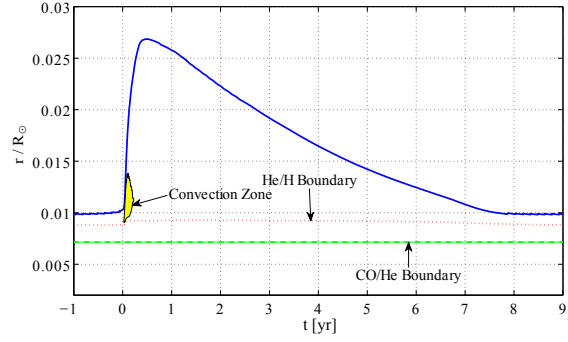}
 \caption{The behaviour of the convection zone during a hydrogen flash. The broken line represents the temporal evolution of the C-O/He boundary.  The dotted line describes the behaviour of the He/H boundary. The upper  line is the radius of the outermost shell. }
 \label{fig:convection}
\end{figure}

\begin{figure} 
\centering
\includegraphics[width=\figsize]{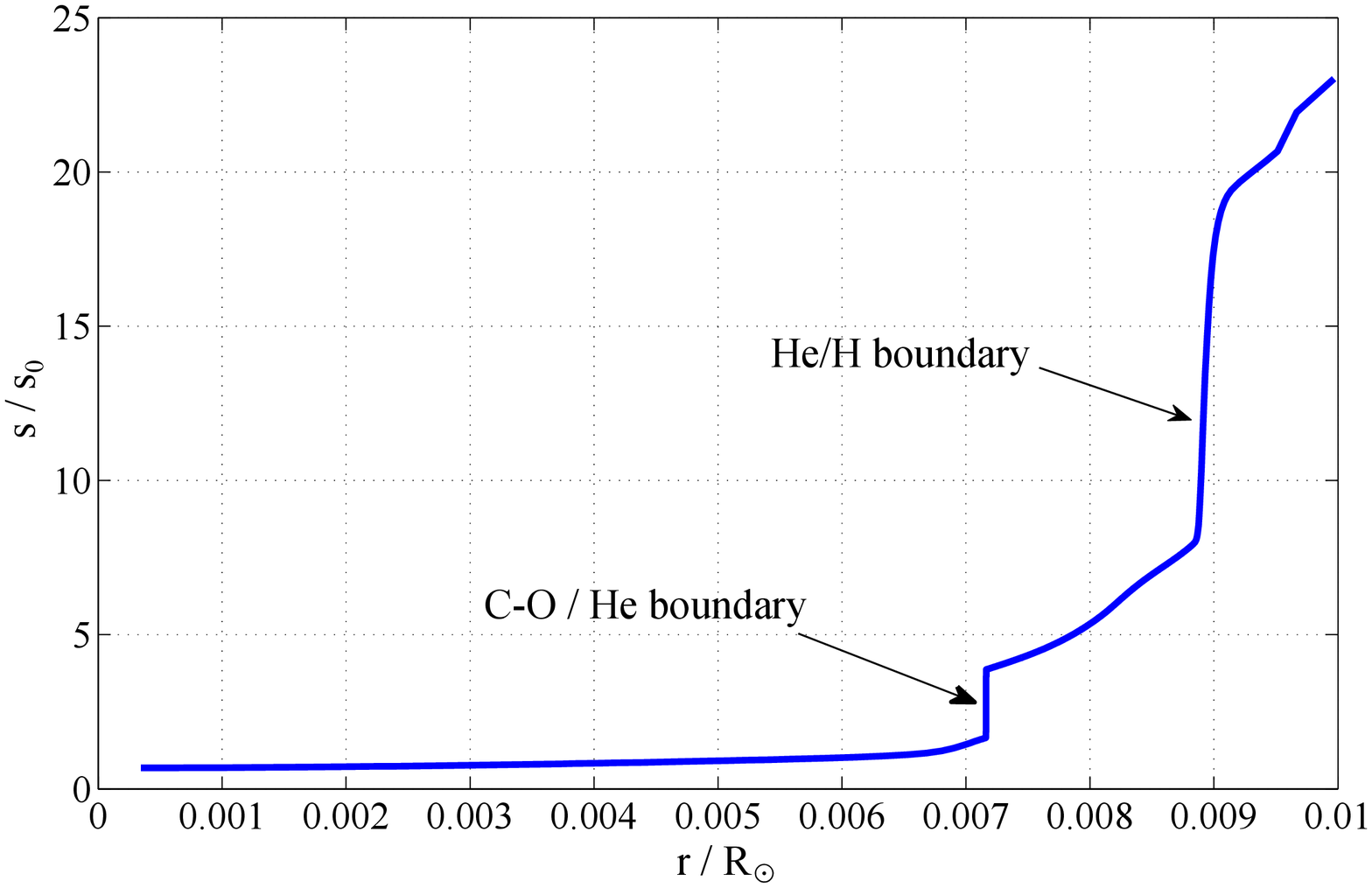}

 \caption{The profile of the specific entropy $s$ at the beginning of the $3000^{th}$ cycle.  The entropy  is given in units of $s_0 \equiv \pi^2\left( M_\odot / N_A\right)^2 \left(G / {\hbar c}\right)^3 K^{-1}$.}
 \label{fig:entropy}
\end{figure}

The history of the convective zone is depicted in fig.~\ref{fig:convection} along with the borders between the hydrogen and helium layers, and between the core and the helium layer. In fig.~\ref{fig:entropy}, the profile of the specific entropy is shown at the beginning of the $3000^{th}$ cycle (see  fig.~\ref{fig:convection}).   Note the the entropy profile is given only for the inner region where the convection did not penetrate. As is obvious, there are two entropy barriers which prevent the convection from penetrating any deeper and mix the C and O extensively over the envelope. The resulting abundances are shown in fig.~\ref{fig:abundances}. 

\subsection{Accretion rate of $10^{-7} M_\odot yr^{-1}$}

According to \cite{Yaron2005}, WDs should erode for accretion rates lower than about $10^{-8} M_\odot yr^{-1}$. However, as we have seen above, the option of achieving the Chandrasekhar limit is also excluded for an accretion rate of $10^{-6}M_\odot yr^{-1}$, because of the strong helium flashes that take place after several thousand hydrogen ignition cycles. We have therefore decided to include the simulation of WDs accreting at an intermediate rate of  $10^{-7}M_\odot yr^{-1}$. We found no helium flashes. Moreover, since the conditions do not satisfy eq.~\ref{eq:expul}, even if there were, there would have been no expulsion.  The results of these models are summarised in table \ref{tab:summary}.
\begin{table*}
\caption{Characteristics of the Nova Envelope for Accretion Rate of $10^{-7} M_\odot yr^{-1}$ }
\begin{center}
\begin{tabular}{|c|c|c|c|c|c|c|}
\hline
$M_{wd} $&Core Temp.& Accreted  Mass & Ejected  Mass & Residual Mass  & Recurrence \cr
 $[M_\odot]$ &   $[10^6K]$ & Per Cycle $[M_\odot]$  &  Per Cycle $[M_\odot]$ &  Per Cycle $[M_\odot]$ & Time [yr] \cr
 \hline
1.00  & 60 & $9\times 10^{-6}$     & $8\times 10^{-6}$    &$1\times 10^{-6}$   &100 \cr
1.25  & 50 & $1.5\times 10^{-6}$  & $1.2\times 10^{-6}$  & $3\times 10^{-7}$  & 16\cr
1.35  & 50  & $3.5\times 10^{-7}$  & $2.6\times 10^{-7}$  & $9\times 10^{-8}$  & 3.7\cr
1.40  & 60 & $5\times 10^{-8}$     & $4\times 10^{-8}$     &$1\times 10^{-8}$   &  0.5\cr
\hline
\end{tabular}
\end{center}
\label{tab:summary}
\end{table*}%

Our results confirm those of \cite{Yaron2005}. Namely, the WD experiences a net  increase in mass when accreting at a rate of $10^{-7} M_\odot yr^{-1}$. However,  the residual mass fraction left on the WD is  relatively small, of order 10-25\%.  Although attempts to identify such a system with known observed ones failed, such systems if existed could in principle approach the Chandrasekhar mass.

\section{Conclusions}
\label{sec:conclusions}

High rate accretion of solar composition material onto WDs prevents degeneracy. According to general wisdom, this should give rise to quiet steady state burning in which the helium ash from the hydrogen burning in the envelope should accrete onto the core, and secularly increase its mass. 

Since eruptions are expected to disrupt the accretion process, our first conclusion is that instead of a steady state process, the accretion and burning take place in eruptive cycles.  
However, this cyclic behaviour is different from classical nova eruptions. We find that the accreted material is only moderately degenerate before it ignites. And once it does, it burns completely. Therefore, any ejecta will contain no detectable hydrogen. Clearly then, the non-existence of hydrogen in observed ejecta does not immediately imply that the accreted matter did not contain any hydrogen as well. Since the lack of observable hydrogen in ejecta is generally considered as one of the strongest evidence suggesting the accretion of pure helium, such claims should be considered more cautiously.   

The last, but most important conclusion we have reached is that high accretion rates, of  $10^{-6}M_{\odot}$, do not lead to the a secular increase in the mass of the white dwarf, because giant helium eruptions take place after a sufficient amount of helium accumulates. These eruptions expel all the accreted layer and even a small amount of the underlying WD. 

We find that helium flash explosions take place for WDs with masses up to 1.35$M_\odot$. From eq.~\ref{eq:expul}, however, there should be no upper limit above which helium cannot expel the material---although more compact systems would require more energy to expel material out of them, they also release more energy during the accretion. 

The above accretion rate is very high and can be maintained by only few systems. The results for a lower accretion rate, of $10^{-7}M_{\odot}$ is qualitatively different. There are no giant helium flashes, and there is a net increase of the WD mass. However, because the mass accretion efficiency is small, it requires the donor to fine tune the mass transfer over a large integrated total mass. It therefore seems unlikely that large increases in the WD mass are possible.   

The above conclusions depend on the energy release by the hydrogen burning. Thus, systems in which helium is accreted will not have their envelope bloated, and any helium eruption will not lead to large mass loss. This possibility is presently under investigation. 

Another important point that is the subject of further investigation is the effects that will arise once super-Eddington conditions are allowed to develop \citep{ShavivNovae}. This should modify the above conclusion for two main reasons. On one hand, the super-Eddington luminosities drive mass loss through continuum driven winds. On the other hand, a higher luminosity will inhibit the atmosphere from bloating, such that the helium flashes will not necessarily be able to drive the envelope away. Thus, it is not a priori clear how the above picture will be modified.

 \section*{Acknowledgements}

This research has been supported by the Israel Science Foundation, grant 1589/10.

  \def\aj{Astron.\ J.}
\def\apj{Ap.\ J.}
\def\apjs{Ap.\ J.\ Supp.}
\def\apjl{Ap.\ J.\ Lett.}
\def\mnras{Mon.\ Not.\ Roy.\ Astro.\ Soc.}
\def\aap{Astron.\ Astrophys.}
\def\araa{Ann.\ Rev.\ Astron.\ Astrophys.}
\def\pasj{Pub.\ Astron.\ Soc.\ Japan}
\def\apss{Astrophys.\ Sp.\ Sci.}
\def\nar{N.\ Astron. Rev.}
\def\nat{Nature}
 

\end{document}